# The Photon and Magnetic Monopoles.

**Eleonora A Mihul and Alexandru C Mihul**
Dep. of Theor. Phys.
Bucharest. University

1. **Time and Lorentz metric.**

Conceptually, all the results are relying on the fact that *time* is not a physically meaning compatible with a *space structured by Lorentz metric.*
    The constant c interpreted as speed is strictly related to the traditional way of measuring it, as length over time.
    The component $x^0$ of a 4-vector $x = (x^0, x^1, x^2, x^3)$ in X, plays a very clue roles in the physical laws. It is not related to time, it is related to the *energy* of the system. The interpretation of $x^0$ as being related to time just by relabeling it by cT and declaring that T represents what we call *time* is physically meaningless [ 1 ].
    We avoid the critic We tried getting a constructive "road" as replica to the R. Penrose book, '"The road to Reality" [ 2 ].
    Let us cite some texts referring to the issue; "In my opinion, the theory of special relativity was not yet complete", {pg.406}.and the most competent and honest critic is the following: "the combined requirements of special relativity and quantum theory forced us into the quagmire of the quantum theory of fields"(pg.870).
    The precise answer is that there are not two theories, special relativity; constant c, and quantum mechanics, constant h. A space X endowed with Lorentz interval **implies** only one constant. We call it **Lorentz constant L = ch** whose dimension is energy multiplies by length.
    For that, besides the structure of X, it has to be defined the concept of energy, of a system represented in X [ 3 ].
    Important is that Lorentz tensor metric being not singular (in "Newtonian space-time" the metric tensor is singular), the dual metric space X* provides a new 4-dimensinal space whose vectors we may interpret as representing what we call 4-vector energy.
    Thus the *4- dimensional energy vector p in X*as the components of the bilinear form*

$$p(x) = p_0 x^0 - \mathbf{p}.\mathbf{x}$$

is well defined. The space X* is just the metric dual space of X. in which a vector $p = (p_0, p_1, p_2, p_3,)$.

Being a metric dual space of X, the Lorentz metric defined in X induces the metric in X*, and as well the dual group of symmetry.

We emphasize that it is obvious that only the simultaneous use of the both spaces lead to the physical interpretable meaning.

For that we have to attach to the component of x and to the components of p measurable entities. It is clear that all components of a vector x we interpret as length and the all components of p we interpret as energy.

Two spaces X and X* play the complementary role in any consequence of the above setting through the functional p(x) = ch [ 3 ].

We proved that we actually represent physical system on a generalized sequence in the field of real numbers, defined on the Lorentz Ordered set in a partial ordered space, directed [ 4 ] with respect to Lorentz ordered relation.

In the old interpretation, causality is a sinecvanon condition of any physical meaningful results emphasized also in private discussions, in particular by A. Wightman.

Therefore we started to find out what are its implications. The first results, "Consequence of Causality" [ 5 ], was performed at the University Galileo Galilee. As in the point of fact, since causality [ 6 ] minces, Ordering on time, we realized that is important the concept of "Ordering", representing in fact our inductive logic, and not necessary "on time".

Therefore we can *eliminate time* and instead of the causality, we are relying on the *Lorentz Ordering*.

Two points $x' = (x'^0, x'^1, x'^2, x'^3,) = (x'^0, \mathbf{x'})$ and $x'' = (x''^0, \mathbf{x''})$ are Lorentz ordered if and only if the interval

$$\text{Int.}(x'-x'') = (x'^0 - x''^0)^2 - (\mathbf{x'} - \mathbf{x''})^2 > 0$$

either

$$\text{Int.}(x'-x'') = (x'^0 - x''^0)^2 - (\mathbf{x'} - \mathbf{x''})^2 = 0$$

and

$$x'^0 > x''^0.$$

As we mentioned already, coherent starting point is to consider that **all components** of the four dimensional real vector space X, structured by **Lorentz** interval are lengths inclusive $x^0$

The purpose of the paper [ 1 ] was to present conceptual content and consequences of the issue. It implies not only the base of "relativity" but also of "Quantum mechanics".

Precisely, it implies the Lorentz Constant L = ch and so we proved, without any other assumption but *Lorentz Ordering* the **Energy - Length Rule** that says:

For an isolated physical system defined by any given 4-dimensional energy vector p in the dual space X*, with $p^2 = 0$, and for $p^2 > 0$ there is an interval $\chi$ in X so that their product is the Lorentz ordering constant L It means $p(\chi) = L$ [ 7 ]v.

Then we were looking actually, for those properties of isolated physical systems implied [ 8 ] by this Energy-Length Rule.

For rest (massive) system, for $\chi = \mathbf{0}$, zero component $\chi^0$ is just the Compton wavelength, whereas for restless photon, $\chi^0$ is its wavelength. The relation $p_0 \chi^0 = ch$ is just the Planck relation of energy. We note that relation $p_0 \chi^0 = ch$ is formally the same for two cases.

**Note**. The Lorentz Constant **L** is a fundamental constant including the constant **h**, as claimed experimentally and expressed in the Planck's formula. Thus Planck launched quantum revolution and so the quantum of radiation-the photon which provides the particle description of light that travels with "velocity " c.

So important is that we proved that the constant, ch, is implied by the Lorentz ordering constraints imposed by the way we are looking to outside world without any other assumption but Lorentz Ordering.

Now for restless systems **[ 9 ]** only the relative orientation between **p** and **x** is left to be chosen**.** There are two invariant possibilities; **p $\perp$ x, photon, and** with the **p parallel** pointing in the same direction with **x , neutrino** and so the **antiparallel case**, **antineutrino, p** pointing in the opposite direction with **x.**
We call them respectively, Transversal and Longitudinal.

For **longitudinal** systems we get $P_0 \chi^0 = \pm 1/2 hc$. The $\chi^0$ is positive for neutrino and negative for antineutrino [8 ]..

**2. Lorentz Ordering.**

Obviously that eliminating the **time** we cannot define **causality** as <u>Ordering</u> on **time**. According to many physicists a result which does not include causality can not have any physical meaning We showed that it is actually <u>Lorentz</u> <u>Ordering as a fundamental assertion.</u>

However the conscious mentality is unavoidable issue in our understanding of physical reality. Inductive logic endowed us with the "ordering tool" with which we quest for understanding of physical reality. This makes us to assert that only the Lorentz **ordered events in X are physically meaningful** Hence, we substitute the time ordering related events what we are accustomed to call causally **related**, by **Lorentz Ordered related events**

We emphasized the enormous consequences of **Lorentz Ordering [3].** The essential idea rests in the requirement that it should be in accordance with the natural world - outside. Therefore we must avoid human artifice or our desire.

The our purpose is to start from the **Book of Nature** and turn to the **Book of Nature for** verifying if our findings are "indicating" in the book".

For **transversal** case, the relation $P_0 \chi^0 - \mathbf{p}\chi = ch$ **means:**
on one side $P_0 \chi^0 = \pm ch$, on the other side the sine of the **p $\perp$ x, representing the helicity of the photon ,** is positive and negative.

The sine positive and respectively negative correspond to the sign of $\chi^0$, are referring to the two cones which define respectively *system and antisystem*. Lorentz ordering and so its consequences are equally for the both cones .

Obviously **the energy is positive** for both of them, it means independent of being it system or its " antisystem" .

Another property imposed by transversally is the keeping **p orthogonal to x [ 11 ] , [ 12]** correspond to right and left handed circular polarization of the photon called sometimes "orbital momentum.

Interesting that the Topology of Minkowski Space **[ 13 ] lays** in the two symptomatic properties; space must be at least 4- dimensional and Euclidean topology is implied for the 3-dimensional space.[ 11 ]
It seems that U(1 ) symmetry of the photon is implicated by the constraint of the symptomatic condition   of the preserving orthogonal structure of 3-dimensional **x** and **p** (orbital momentum)**.**

**So photon looks like a "Spinning, rotating, sliding** systems" so that at each point of a ray, roughly speaking, interpreted mechanically, a "circle "is accomplished" so that the end of which is the start of the circle in the adjacent point.

It looks like represented by a cylindrical space topology **[12 ]**. The stability of the photon is owing to the infinite connected topology. Then the winding number defined in the first group of homotopy, indicates uniquely the length we measure of traveling photon, which does not know what the REST means.

Interesting that the Topology of Minkowski Space **[12] lay** in the two symptomatic properties; space must be at least 4- dimensional and Euclidean topology is implied for the 3-dimensional space.

Actually we deal in each point with a **2-dimensional membrane** which must manifest itself  by a property which obviously can not be interpreted using mechanic concept like "orbital momentum". We interpret it to of a pure magnetic nature .

The effect of this pure magnetic property is manifested in the photon Hall Effect [.14 ].

We should not call it magnetic charge, since it seems that any charge is carried by a rest system. Instead these pure magnetic properties are related to a restless system, the photon.

Hence it rather seems that what we are tempting to call "positive and negative" magnetic charge are actually the representation of right and left handed circular polarizations of the photon [ 1 ]. Thereby the effect of this "purely magnetic"2-dimensional membrane is manifested like positive respectively negative magnetic monopole.

 Being related to the sign positive and respectively negative of $\chi^0$, corresponding to the two cones plays the role of system and antisystem .So the two positive and negative monopoles are not independent, in the sense that one is antisystem of the other, as it was expecting.

It seems that we arrived at a new structure of the photon, as a consequence of the Energy-Length Rule.   Does it corresponds  to the wish of many generations of experimental and theoretical physicists, to discover  so called Magnetic Monopoles The  theoretical and experimental struggle to find the magnetic monopole is so well described in [ 15 ] what is completely ignored in our discussion.

In our approach one has to find out experimentally how the photon Hall effect is dependent of:

1.the use of the two diverse right and left handed circular polarizations of the photon,
**2.** the energy of the photon ,is expecting to be the major property.
**3.** the intensity (number) of the photon.

**3. Reflections.**

Important is that the properties of isolated photons and neutrinos, as well as magnetic monopoles are consequences of the above structure without any other assumption, but Lorentz ordering constraint.

It implies the interpretation of $x^0$ as length **[ 14 ]** having no connection at all with the concept of time. We do not use any theoretical or experimental indication presented so complete in [15]. The road we take is simple.

Obviously that causality [16] played a primary condition in the way we interpreted outside world in the mechanical concepts. The point is that for interpreting experimental information in mechanical way, it seems meaningless to introduce like Schwarzschild metric, for instance, and prove that it is equivalent to a certain classical potential [17].

I should like to mention that the first time I presented the relation between causality and Planck constant was at 1989 [18].

4. **Conclusion**

The Lorentz Ordering is the principle of particle physics. The component $x^0$ is the hart of it.

The two signs of $x^0$ define the two cones. Since the Lorentz Ordering is equally valid in both of them, for any system defined by a sign of $x^0$ there is its antisystem defined by opposite sign. Is this manifested, for instance, in the electron-positron pair production by $\Gamma =$ photon, that actually "absorbs" the two monopoles?

It looks like indeed the "conservation" of system and antisystem imposed by Lorentz metric is as powerful as the conservation of the energy that also is a consequence of Lorentz metric [ 5 ].

5. **Acknowledgement,**

We would like to thank Milla-Baldo-Ceolin and the physicists of her department for many useful suggestions and also to the staff of theoretical department of J.I N R, Dubna.

6. **References**.